\documentclass[a4paper,10pt]{article}
\usepackage{latexsym}
\usepackage{amssymb} %
\title{Some Aspects of Quantum Entanglement for  CAR Systems} 
\author{Hajime Moriya}
\date{}
\begin{document}
\maketitle
\begin{abstract}
We show   some  distinct   features of  quantum entanglement for bipartite 
 CAR systems such as the failure  of triangle inequality
 of von Neumann entropy and 
 the  possible change of our entanglement degree under {{\it local}} 
operations.
Those are due to the nonindependence 
 of CAR systems and never occur  in any algebraic independent systems.
We introduce a new notion {{\it half-sided entanglement}}.
\end{abstract}
\newtheorem{thm}{Theorem}%
\newtheorem{df}{Definition}%
\newtheorem{col}[thm]{Corollary}%
\newtheorem{lem}[thm]{Lemma}%
\newtheorem{pro}[thm]{Proposition}%
\renewcommand{\theparagraph}{\Alph{paragraph}}%
\newtheorem{rem}{\it{Remark}}
\setcounter{secnumdepth}{5}%
%
\newcommand{\cstar}{{\bf C}^{\ast}}%
\newcommand{\wstar}{{\bf W}^{\ast}}%
\newcommand{\R} {{\mathbb{R}}}%
\newcommand{\Z}{{\mathbb{Z}}}%
\newcommand{\C} {{\mathbb{C}}}%
\newcommand{\NN}{{\mathbb{N}}}%
\newcommand{\qedb}{\hbox{\rule[-2pt]{3pt}{6pt}}}%
\newcommand{\Al}{{\cal A}}%
\newcommand{\Bl}{{\cal B}}%
\newcommand{\Cl}{{\cal C}}%
\newcommand{\Clbar}{\overline{\Cl} }%
\newcommand{\nonum}{\nonumber}%
\newcommand{\Mat}{{\mathrm  M}_{2}(\C)}%
\newcommand{\Matfour}{{\mathrm M}_{4}(\C)}%
\newcommand{\Tr}{\mathbf{Tr}}%
\newcommand{\Tri}{\mathbf{Tr}_{\rm{i} } }%
\newcommand{\Trii}{\mathbf{Tr}_{\rm{ii} } }%
\newcommand{\TrA}{{\mathbf{Tr}}_{A} }%
\newcommand{\TrB}{{\mathbf{Tr}}_{B} }%
\newcommand{\tauA}{\tau_{A}}%
\newcommand{\tauB}{\tau_{B}}%
\newcommand{\aicr}{a_i^{\ast}}%
\newcommand{\ai}{a_i}%
\newcommand{\ajcr}{a_j^{\ast}}%
\newcommand{\aj}{a_j}%
\newcommand{\afcr}{a_1^{\ast}}%
\newcommand{\af}{a_1}%
\newcommand{\ascr}{a_2^{\ast}}%
\newcommand{\as}{a_{2}}%
\newcommand{\bicr}{b_i^{\ast}}%
\newcommand{\bi}{b_i}%
\newcommand{\bjcr}{b_j^{\ast}}%
\newcommand{\bj}{b_j}%
\newcommand{\bfcr}{b_1^{\ast}}%
\newcommand{\bff}{b_1}%
\newcommand{\bscr}{b_2^{\ast}}%
\newcommand{\bs}{b_{2}}%
\newcommand{\Alf}{\Al_{1}^{\rm{car}}}%
\newcommand{\Als}{\Al_{2}^{\rm{car}}}%
\newcommand{\Ali}{\Al_{i}^{\rm{car}}}%
\newcommand{\Alwhole}{\Al_{1,2}^{\rm{car}}}%
\newcommand{\Alsspin}{\Al_{2}^{\rm{spin}}}%
\newcommand{\Alfspin}{\Al_{1}^{\rm{spin}}}%
\newcommand{\Alispin}{\Al_{i}^{\rm{spin}}}%
\newcommand{\Alwholeeven}{  \Al_{1,2,\,+}^{\rm{car}} }%
\newcommand{\Alwholeodd}{  \Al_{1,2,\,-}^{\rm{car}} }%
\newcommand{\Alfeven}{ \Al_{1,\,+}^{\rm{car}}  }%
\newcommand{\Alfodd}{ \Al_{1,\,-}^{\rm{car}}  }    %
\newcommand{\Alseven}{ \Al_{2,\,+}^{\rm{car}}  }%
\newcommand{\Alsodd}{\Al_{2,\,-}^{\rm{car}}} %
\newcommand{\Alieven}{ \Al_{i,\,+}^{\rm{car}}  }%
\newcommand{\Aliodd}{\Al_{i,\,-}^{\rm{car}}} %
\newcommand{\sumij}{\sum_{i,j}}%
\newcommand{\identitybf}{{\mathbf{1} } }
\newcommand{\proofend}{{\hfill $\square$}\  \\}
\newcommand{\efij}{e^{1}_{(i,j)}}%
\newcommand{\efji}{e^{1}_{(j,i)}}%
\newcommand{\efii}{e^{1}_{(i,i)}}%
\newcommand{\efjj}{e^{1}_{(j,j)}}%
\newcommand{\efoo}{e^{1}_{(1,1)}}%
\newcommand{\efot}{e^{1}_{(1,2)}}%
\newcommand{\efto}{e^{1}_{(2,1)}}%
\newcommand{\eftt}{e^{1}_{(2,2)}}%
\newcommand{\esij}{e^{2}_{(i,j)}}%
\newcommand{\esji}{e^{2}_{(j,i)}}%
\newcommand{\esii}{e^{2}_{(i,i)}}%
\newcommand{\esjj}{e^{2}_{(j,j)}}%
\newcommand{\esoo}{e^{2}_{(1,1)}}%
\newcommand{\esot}{e^{2}_{(1,2)}}%
\newcommand{\esto}{e^{2}_{(2,1)}}%
\newcommand{\estt}{e^{2}_{(2,2)}}%
\newcommand{\esspinij}{e^{2({\rm{spin}})}_{(i,j)}}%
\newcommand{\esspinji}{e^{2({\rm{spin}})}_{(j,i)}}%
\newcommand{\esspinii}{e^{2({\rm{spin}})}_{(i,i)}}%
\newcommand{\esspinjj}{e^{2({\rm{spin}})}_{(j,j)}}%
\newcommand{\esspinoo}{e^{2({\rm{spin}})}_{(1,1)}}%
\newcommand{\esspinot}{e^{2({\rm{spin}})}_{(1,2)}}%
\newcommand{\esspinto}{e^{2({\rm{spin}})}_{(2,1)}}%
\newcommand{\esspintt}{e^{2({\rm{spin}})}_{(2,2)}}%
\newcommand{\vrho}{\varrho}%
\newcommand{\vrhof}{\varrho_{1}}%
\newcommand{\vrhos}{\varrho_{2}}%
\newcommand{\hrho}{\hat{\rho}}%
\newcommand{\hrhof}{\hrho_{1}}%
\newcommand{\hrhos}{\hrho_{2}}%
\newcommand{\hrhospin}{\hrho_{2}^{{\rm{spin}}}}
\newcommand{\hrhosspin}{\hrho_{2}^{{\rm{spin}}}}%
\newcommand{\hrhofspin}{\hrho_{1}^{{\rm{spin}}}}%
\newcommand{\trho}{\tilde{\rho}}%
\newcommand{\trhof}{\trho_{1}}%
\newcommand{\trhos}{\trho_{2}}%
\newcommand{\brho}{\bar{\rho}}%
\newcommand{\brhof}{\brho_{1}}%
\newcommand{\brhos}{\brho_{2}}%
\newcommand{\brhoi}{\brho_{i}}%
\newcommand{\brhoj}{\brho_{j}}%
\newcommand{\brhop}{\brho^{\prime}}%
\newcommand{\brhopf}{\brho^{\prime}_{1}}%
\newcommand{\brhops}{\brho^{\prime}_{2}}%
\newcommand{\brhopi}{\brho^{\prime}_{i}}%
\newcommand{\brhopj}{\brho^{\prime}_{j}}%
\newcommand{\vrhoDECOMPf}{\varrho^{\{1\}}}%
\newcommand{\vrhoDECOMPs}{\varrho^{\{2\}}}%
\newcommand{\vrhoi}{\vrho^{\{i\}}}%
\newcommand{\vrhospin}{\varrho_{2}^{{\rm{spin}}}}
\newcommand{\vrhosspin}{\varrho_{2}^{{\rm{spin}}}}%
\newcommand{\vrhofspin}{\varrho_{1}^{{\rm{spin}}}}%
\newcommand{\Evrhospm}{\vrhos^{\pm}(\vpp,\,\vp)}%
\newcommand{\Evrhosp}{\vrhos^{+}(\vpp,\,\vp)}%
\newcommand{\Evrhosm}{\vrhos^{-}(\vpp,\,\vp)}%
\newcommand{\Hil}{{\cal H}}%
\newcommand{\Hilf}{{\cal H}_{1}}%
\newcommand{\Hils}{{\cal H}_{2}}%
\newcommand{\HilA}{{\cal H}_{A}}%
\newcommand{\HilB}{{\cal H}_{B }}%
\newcommand{\HilAB}{\HilA \otimes \HilB}%
\newcommand{\zetai}{\zeta_{i}}%
\newcommand{\zetaA}{\zeta^{A}   }%
\newcommand{\zetaB}{\zeta^{B}  }%
\newcommand{\zetaAj}{\zeta^{A}_{(j)}}%
\newcommand{\zetaBj}{\zeta^{B}_{(j)}}%
\newcommand{\zetaBk}{\zeta^{B}_{(k)}}%
\newcommand{\xif}{\xi^{1} }%
\newcommand{\xis}{\xi^{2} }%
\newcommand{\xifi}{\xi^{1}_{i} }%
\newcommand{\xifj}{\xi^{1}_{j} }%
\newcommand{\xifo}{\xi^{1}_{1} }%
\newcommand{\xift}{\xi^{1}_{2} }%
\newcommand{\xisi}{\xi^{2}_{i} }%
\newcommand{\xisj}{\xi^{2}_{j} }%
\newcommand{\xiso}{\xi^{2}_{1} }%
\newcommand{\xist}{\xi^{2}_{2} }%
\newcommand{\xiij}{\xi_{i,j}  }%
\newcommand{\xioo}{\xi_{1,1}  }%
\newcommand{\xiot}{\xi_{1,2}  }%
\newcommand{\xito}{\xi_{2,1}  }%
\newcommand{\xitt}{\xi_{2,2}  }%
\newcommand{\Txiij}{\xifi \otimes \xisj  }%
\newcommand{\Txiii}{\xifi \otimes \xisi  }%
\newcommand{\Txioo}{\xifo \otimes \xiso}%
\newcommand{\Txiot}{\xifo \otimes \xist }%
\newcommand{\Txito}{\xift \otimes \xiso}%
\newcommand{\Txitt}{\xift \otimes \xist }%
\newcommand{\PfoTxioo}{\\Pxifo xifo \otimes \xiso}%
\newcommand{\PfoTxitt}{\Pxifo xift \otimes \xist }%
\newcommand{\PftTxioo}{\\Pxift xifo \otimes \xiso}%
\newcommand{\PftTxitt}{\Pxift xift \otimes \xist }%
\newcommand{\PsoTxioo}{\xifo \otimes \Pxiso \xiso}%
\newcommand{\PsoTxitt}{\xift \otimes \Pxiso \xist }%
\newcommand{\PstTxioo}{\xifo \otimes \Pxist \xiso}%
\newcommand{\PstTxitt}{\xift \otimes \Pxist \xist }%
\newcommand{\vp}{\varphi}%
\newcommand{\ome}{\omega}%
\newcommand{\omef}{\ome_{1}}%
\newcommand{\omes}{\ome_{2}}%
\newcommand{\omefi}{\ome_{1,i}}%
\newcommand{\omesi}{\ome_{2,i}}%
\newcommand{\omeA}{\ome_{\Al}}%
\newcommand{\omeB}{\ome_{\Bl}}%
\newcommand{\omefj}{\ome_{1,j}}%
\newcommand{\omesj}{\ome_{2,j}}%
\newcommand{\ik}{i_{k} }%
\newcommand{\omei}{\ome_{i}}%
\newcommand{\omej}{\ome_{j}}%
\newcommand{\omek}{\ome_{k}}%
\newcommand{\omeik}{\ome_{\ik}}%
\newcommand{\vpf}{\vp_{1}}%
\newcommand{\vps}{\vp_{2}}%
\newcommand{\vpp}{\vp^{\prime}}%
\newcommand{\psif}{\psi_{1}}%
\newcommand{\psis}{\psi_{2}}%
\newcommand{\phif}{\phi_{1}}%
\newcommand{\phis}{\phi_{2}}%
\newcommand{\phiA}{\phi_{A}}%
\newcommand{\phiB}{\phi_{B}}%
\newcommand{\etavp}{\eta_{(\vp)}}%
\newcommand{\etaome}{\eta_{(\ome)}}%
\newcommand{\omeeta}{\ome_{\eta}}%
\newcommand{\omexifi}{\ome_{\xifi}}%
\newcommand{\etaomepara}{\eta_{(\ome)}^{\{ \thetaoo \} } }%
\newcommand{\etabrho}{\eta_{(\brho)}}%
\newcommand{\vpeta}{\vp_{\eta}}%
\newcommand{\vpetavp}{\vp_{\etavp}}%
\newcommand{\vpetaome}{\vp_{\etaome}}%
\newcommand{\iso}{\cong}%
\newcommand{\Uf}{U_{1}}%
\newcommand{\Us}{U_{2}}%
\newcommand{\Ui}{U_{i}}%
\newcommand{\Uj}{U_{j}}%
\newcommand{\CONS}{\bigl\{ \xiij  \bigr\}}%
\newcommand{\ccij}{c_{i,j}}%
\newcommand{\ccoo}{c_{1,1}}%
\newcommand{\ccot}{c_{1,2}}%
\newcommand{\ccto}{c_{2,1}}%
\newcommand{\cctt}{c_{2,2}}%
\newcommand{\ccomeij}{c{(\ome)}_{i,j}}%
\newcommand{\ccomeoo}{c{(\ome)}_{1,1}}%
\newcommand{\ccomeot}{c{(\ome)}_{1,2}}%
\newcommand{\ccometo}{c{(\ome)}_{2,1}}%
\newcommand{\ccomett}{c{(\ome)}_{2,2}}%
%
%
%
\newcommand{\MatCARfnophase}[1]
 {\left(
  \begin{array}{cc}
   \cos^{2}(#1)& \cos(#1) \sin(#1) \\ 
   \cos(#1) \sin(#1) & \sin^{2}(#1)
 \end{array}
      \right) }%
\newcommand{\MatSPINnophase}[1]
 {\left(
   \begin{array}{cc}
     \cos^{2}(#1)& \cos(#1) \sin(#1) \\ 
     \cos(#1) \sin(#1) & \sin^{2}(#1)
  \end{array}
      \right) }%
\newcommand{\MatCARsnophase}[2]
 {\left(
   \begin{array}{cc}
     \cos^{2}(#1)&   \{\cos^{2}(#2) -\sin^{2}(#2)\}  \left(\cos(#1) \sin(#1)\right) \\ 
   \{\cos^{2}(#2) -\sin^{2}(#2)\} \left( \cos(#1)\sin(#1) \right)& \sin^{2}(#1)
  \end{array}
      \right) }%
%
\newcommand{\MatCARf}[2]
 {\left(
  \begin{array}{cc}
   \cos^{2}(#1)& e^{i #2} \cos(#1) \sin(#1) \\ 
   e^{-i #2} \cos(#1) \sin(#1) & \sin^{2}(#1)
 \end{array}
      \right) }%
\newcommand{\MatSPIN}[2]
 {\left(
   \begin{array}{cc}
     \cos^{2}(#1)&   e^{i #2} \cos(#1) \sin(#1) \\ 
      e^{-i #2} \cos(#1) \sin(#1) & \sin^{2}(#1)
  \end{array}
      \right) }%
\newcommand{\MatCARs}[3]
 {\left(
   \begin{array}{cc}
     \cos^{2}(#1)
 &   e^{i #3} \{\cos^{2}(#2) -\sin^{2}(#2)\}  
\left(\cos(#1) \sin(#1)\right) \\ 
  e^{-i #3}  \{\cos^{2}(#2) -\sin^{2}(#2)\} \left( \cos(#1)\sin(#1) \right)
 & \sin^{2}(#1)
  \end{array}
      \right) }%
\newcommand{\MatCARsg}[3]
 {\left(
   \begin{array}{cc}
     \cos^{2}(#1)
 &    g(#2) \cdot e^{i #3} \cos(#1) \sin(#1) \\ 
    g(#2) \cdot  e^{-i #3}   \cos(#1)\sin(#1) 
 & \sin^{2}(#1)
  \end{array}
      \right) }%
\newcommand{\MatCARdiag}[1]
 { 
\left( \begin{array}{cc}
     \cos^{2}(#1) &0 \\ 
    0 & \sin^{2}(#1)
  \end{array}
      \right) 
  }%
\newcommand{\MatCARdiagoo}[1]
 { 
\left( \begin{array}{cc}
     \cos^{2}(#1) &0 \\ 
    0 & 0
  \end{array}
      \right) 
  }%
\newcommand{\MatCARdiagtt}[1]
 { 
\left( \begin{array}{cc}
     0 &0 \\ 
    0 & \sin^{2}(#1) 
  \end{array}
      \right) 
  }%
\newcommand{\MatNR}[4]
 { 
\left( \begin{array}{cc}
     #1 & #2 \\ 
    #3 & #4 
  \end{array}
      \right) 
  }%
\newcommand{\Hill}{\Hil^{l}}%
\newcommand{\Hilr}{\Hil^{r}}%
\newcommand{\Hillr}{\Hill \otimes \Hilr }%
\newcommand{\LL}{{\cal L}}%
\newcommand{\etaf}{\eta^{1}}%
\newcommand{\etas}{\eta^{2}}%
\newcommand{\etafome}{\eta^{1}_{(\ome)}}%
\newcommand{\etasome}{\eta^{2}_{(\ome)}}%
\newcommand{\etafomepara}{\eta^{1,\{\thetaoo\} }_{(\ome)}    }%
\newcommand{\etasomepara}{\eta^{2,\{\thetaoo\} }_{(\ome)}    }%
%
\newcommand{\etavrho}{\eta_{(\vrho)}}
\newcommand{\etafvrho}{\eta^{1}_{(\vrho)}}%
\newcommand{\etasvrho}{\eta^{2}_{(\vrho)}}%
\newcommand{\omespin}{\ome_{2}^{{\rm{spin}}}}%
\newcommand{\thetao}{\theta_{1} }%
\newcommand{\thetat}{\theta_{2} }%
\newcommand{\thetap}{\theta^{\prime}}%
\newcommand{\thetapo}{\theta^{\prime}_{1}}%
\newcommand{\thetapt}{\theta^{\prime}_{2}}%
\newcommand{\thetaoo}{\theta_{11} }%
\newcommand{\thetaot}{\theta_{12} }%
\newcommand{\thetato}{\theta_{21} }%
\newcommand{\thetatt}{\theta_{22} }%
\newcommand{\phip}{\phi^{\prime}}%
\newcommand{\cosphi}{\cos (\phi)}%
\newcommand{\cosphip}{\cos (\phip)}%
\newcommand{\sinphi}{\sin (\phi)}%
\newcommand{\sinphip}{\sin (\phip)}%
\newcommand{\cosphif}{\cos (\phif)}%
\newcommand{\sinphif}{\sin (\phif)}%
\newcommand{\cosphis}{\cos (\phis)}%
\newcommand{\sinphis}{\sin (\phis)}%
\newcommand{\cosphiA}{\cos (\phiA)}%
\newcommand{\sinphiA}{\sin (\phiA)}%
\newcommand{\cosphiB}{\cos (\phiB)}%
\newcommand{\sinphiB}{\sin (\phiB)}%
\newcommand{\cosphisq}{\cos^{2}(\phi)}%
\newcommand{\sinphisq}{\sin^{2}(\phi)}%
\newcommand{\cosphipsq}{\cos^{2}(\phip)}%
\newcommand{\sinphipsq}{\sin^{2}(\phip)}%
\newcommand{\minusfour}{\!\!\!\!}%
\newcommand{\minussix}{\!\!\!\!\!\!}%
\newcommand{\varpif}{\varpi_{1} }%
\newcommand{\varpis}{\varpi_{2} }%
\newcommand{\vtheta}{\vartheta}%
\newcommand{\vthetap}{\vartheta^{\prime}}%
\newcommand{\vthetaf}{\vartheta_{1}}%
\newcommand{\vthetas}{\vartheta_{2}}%
\newcommand{\Pind}{P_{i}}%
\newcommand{\Pinda}{P_{\alpha(i)}}%
\newcommand{\lam}{\lambda}%
\newcommand{\lami}{\lambda_{i} }%
\newcommand{\lamij}{\lambda_{i,j} }%
\newcommand{\lamj}{\lambda_{j} }%
\newcommand{\lamk}{\lambda_{k} }%
\newcommand{\lamik}{\lambda_{\ik} }%
\newcommand{\PdagiPi}{ \Pind^{\ast} \Pind }%
\newcommand{\PdagiPia}{ \Pinda^{\ast} \Pinda }%
\newcommand{\Pxifi}{P(\xifi)}%
\newcommand{\Pxifj}{P(\xifj)}%
\newcommand{\Pxifo}{P(\xifo)}%
\newcommand{\Pxift}{P(\xift)}%
\newcommand{\Pxisi}{P(\xisi)}%
\newcommand{\Pxisj}{P(\xisj)}%
\newcommand{\Pxiso}{P(\xiso)}%
\newcommand{\Pxist}{P(\xist)}%
\newcommand{\PdagiPixi}{ \Pxifi^{\ast} \Pxifi }%
\newcommand{\PdagiPixiS}{ \Pxisi^{\ast} \Pxisi }%
\newcommand{\lamal}{\lambda_{\alpha} }%
\newcommand{\lamali}{\lambda_{\alpha(i)} }%
\newcommand{\ali}{{\alpha(i)} }%
\newcommand{\omeal}{\ome_{\alpha} }%
\newcommand{\omep}{\ome^{\prime} }%
\newcommand{\omeali}{\ome_{\alpha(i)} }%
\newcommand{\Fbar}{\overline{F}}%
\newcommand{\rhobar}{\overline{\rho}}%
\newcommand{\Eol}{\overline{E}}%
\newcommand{\Eul}{\underline{E}}
\newcommand{\FbarBUNKAI}{{ \ome=\sum \lamal \omeal }\atop{ \{ \Pinda \}}   }
\newcommand{\lamvrhoi}{\lam_{ \{ i\}  }}%
\newcommand{\lamvrhof}{\lam_{  1  }(\vp)}%
\newcommand{\lamvrhos}{\lam_{ 2  }(\vp)}%
\newcommand{\Dome}{D_{\ome}}%
\newcommand{\Domef}{D_{\ome_{1}}}%
\newcommand{\Domes}{D_{\ome_{2}}}%
\newcommand{\dome}{{\widehat{d}}_{\ome}}%
\newcommand{\EE}{\cal E}%
\newcommand{\EBS}{\widetilde{\EE}}
\newcommand{\EB}{{\cal E}}
\section{Introduction}
\label{sec:INTRO}
In this Letter, we   investigate  bipartite   
  CAR  systems  from the viewpoint of {\it {quantum entanglement}}.
Quantum entanglement of states refers to  the 
  quantum  correlations  at  separated regions
 which    cannot be reduced to 
the  classical (probability) theory.  
 
A pair of quantum systems  
$\Al$ and $\Bl$  are   given.
Here $\Al$ and $\Bl$ are  the algebras representing 
 quantum subsystems.
 For the usual cases  which have  been studied extensively 
in quantum information theory, 
 these $\Al$ and $\Bl$ 
are   assumed to be algebraically independent, 
that is, $\Al$ and $\Bl$ commute elementwise 
  and the total system $\Cl$   is 
given by   
  the tensor product $\Al\otimes \Bl$.
We are interested in quantum entanglement 
 between  pairs of subsystems 
 which are coupled by 
 different  kind of algebraic relations    other than the 
tensor product.

We treat a bipartite CAR system  which is a typical
 example of  nonindependent systems.
Let $\Alf$ and $\Als$ 
 be  finite-dimensional  CAR systems representing 
 a pair of disjoint 
 subsystems. 
The total system  $\Alwhole$ is given by $\Alf \vee \Als$,
  the algebra 
algebraically  generated by  $\Alf$ and $\Als$.
 For the sake of simplicity, we consider 
a spinless  one-particle Fermion in each  subsystem (i.e., one degree of freedom for each region). 

The  bipartite CAR pair ($\Alf$, $\Als$) are  not 
algebraically  independent. 
Furthermore, we show that ($\Alf$, $\Als$)
 are  not  
 statistically   independent. 
  
We give an  entanglement degree 
 which makes  sense for any 
 pair of finite-dimensional 
 subsystems   $(\Al,\,\Bl)$ of  
 a general $\cstar$-algebra $\Cl$ which 
represents  the total system.
(In  \cite{NARNentangle}, other kind of entanglement degrees  
 are defined 
 for general finite-dimensional algebraically independent pairs $(\Al,\,\Bl)$
sitting in  $\Cl$.)

There is  some difference    on quantum entanglement between 
 the  CAR systems   and  general  tensor-product systems. 
We show that the triangle inequality of von Neumann entropy does not 
 hold for the bipartite CAR systems. 

We introduce a new notion 
 which we call 
{\it{half-sided entanglement}}
 in terms  of 
the asymmetry of marginal entropies. 
 For  tensor-product  systems, it is trivially 0
 for any state, but it can take strictly positive value for 
CAR systems.
 We
 compute  the  degree of the half-sided entanglement  for some states of 
 the bipartite CAR system.
 
 We study how 
 {\it  {local}} operations on a half-sided region  affect the
quantum  entanglement for the bipartite CAR system. 
We show that 
   the local  automorphisms  can change  the entanglement degree in
 the bipartite CAR system
in contrast to the  tensor-product systems.
\section{Preliminaries}
\label{sec:SETTING}
\subsection{Bipartite CAR Systems}
\label{subsec:CAR}
Let $\aicr$ and $\ai$  be  creation 
and  annihilation
 operators, respectively,   satisfying the canonical anticommutation relations 
(CAR):
\begin{eqnarray}
\label{eq:CAR1}
\{ \aicr, \aj \}=\delta_{i,j}\, \identitybf,\quad 
\{ \aicr, \ajcr \}=\{ \ai, \aj \}=0,
\end{eqnarray}
where $\{A, B\}=AB+BA$ (anticommutator),  $i$, $j=$  $1$ or $2$, 
 $\delta_{i,j}=1$ for  $i = j$ and 
  $\delta_{i,j}=0$ for  $i\neq j$.

Let $\Alwhole$ be a $\cstar$- algebra generated by 
$\{ \aicr,\, \ai |\; i=1,2 \}$.
Let $\Alf$ be a $\cstar$-subalgebra of $\Alwhole$
 generated by 
$\afcr$ and $\af$, and 
 $\Als$ be a $\cstar$-subalgebra 
 generated by $\ascr$ and $\as$. 
Each $\Ali$ is imbedded in $\Alwhole$ and $\Alf \vee \Als=\Alwhole$.

We define 
\begin{eqnarray}
\label{eq:Elf}
\efoo  \equiv \afcr \af,  
\quad \efot \equiv \afcr, \quad 
\efto \equiv \af, \quad  \eftt \equiv \af \afcr, 
\end{eqnarray}
 and 
\begin{eqnarray}
\label{eq:Els}
\esoo  \equiv  \ascr \as,\quad  \esot \equiv \ascr, \quad 
\esto \equiv \as,\quad  \estt \equiv \as \ascr.
\end{eqnarray}
Then $\Bigl\{ \efij \Bigr\}_{i,j}$ is a 
system of  matrix units of 
$\Alf \bigl( \iso \Mat \bigr)$,
 and $\Bigl\{ \esij \Bigr\}_{i,j}$  
is that  of $\Als \bigl( \iso \Mat \bigr)$.
Let $\Alsspin$ be a relative commutant of $\Alf$ in $\Alwhole$, that is,  
\begin{eqnarray*}
\label{eq:Commu}
\Alsspin \equiv  \{\Alf\}^{\prime} \cap \Alwhole.
\end{eqnarray*}
 We also define  $\Alfspin \equiv  \{\Als\}^{\prime} \cap \Alwhole$.

The algebraic extension of the 
map 
\begin{eqnarray}
\label{eq:THETA}
\Theta(\aicr)=-\aicr, \quad 
\Theta(\ai)=-\ai\  (i=1,2)
\end{eqnarray}
 is  a $*$-automorphism of $\Alwhole$ 
  and will 
 be denoted by the same symbol $\Theta$.
 The even and odd parts of $\Alwhole$ are given  by
\begin{eqnarray*}
\label{eq:EvenOdd0}
\Alwholeeven \equiv \bigl\{ A \in \Alwhole \, |\, \Theta(A)=A  \bigr\}, \quad
\Alwholeodd  \equiv \bigl\{ A \in \Alwhole \, |\, \Theta(A)=-A  \bigr\}.
\end{eqnarray*}
In the same way, we define 
\begin{eqnarray*}
\label{eq:EvenOdd1}
\Alfeven \equiv \bigl\{ A \in \Alf \, |\, \Theta(A)=A  \bigr\}, \quad
\Alfodd  \equiv \bigl\{ A \in \Alf \, |\, \Theta(A)=-A  \bigr\},
\end{eqnarray*}
and 
\begin{eqnarray*}
\label{eq:EvenOdd2}
\Alseven \equiv \bigl\{ A \in \Als \, |\, \Theta(A)=A  \bigr\}, \quad
\Alsodd  \equiv \bigl\{ A \in \Als \, |\, \Theta(A)=-A  \bigr\}.
\end{eqnarray*}

We  introduce the so-called  Klein-Wigner transformation 
 on $\Als$ as
\begin{eqnarray}
\label{eq:KLEIN1}
 \ascr \mapsto \Uf \ascr \equiv \bscr, \quad  
  \as \mapsto  \Uf  \as \equiv \bs,
\end{eqnarray}
 where $\Uf  \equiv \afcr\af-\af\afcr \bigl( \in \Alf \bigr)$.
 In the same manner, we introduce 
 the Klein-Wigner transformation 
 on $\Alf$ as 
\begin{eqnarray}
\label{eq:KLEIN2}
 \afcr \mapsto \Us \afcr \equiv \bfcr, \quad
  \af \mapsto  \Us  \af \equiv \bff,
\end{eqnarray}
 where $\Us  \equiv \ascr\as-\as\ascr \bigl( \in \Als \bigr)$.

Obviously,  $\Ui=\Ui^{\ast}$, $\Ui \Ui^{\ast}=\identitybf$,
and  $\Ui \in \Alieven$.   
 It follows    from (\ref{eq:CAR1}) 
that $\bicr$ and $\bi$ satisfy the CAR:
 \begin{eqnarray}
\label{eq:CARb}
\{ \bicr, \bj \}=\delta_{i,j}\, \identitybf, \quad
\{ \bicr, \bjcr \}=\{ \bi, \bj \}=0 \quad (i,j=1, 2).  
\end{eqnarray}
It is easy to see that 
$\Alispin$ is algebraically generated by $\bicr$ and $\bi$,
 and  is isomorphic to $\Ali$ by (\ref{eq:KLEIN2})
 and (\ref{eq:KLEIN1}) for 
 $i=1,\,2$. 
We have   
\begin{eqnarray*}
\label{eq:CAR5}
\Alwhole=\Alf \otimes \Alsspin=\Alfspin \otimes \Als.
\end{eqnarray*}

From now on, we shall be mainly concerned with the former tensor 
 product structure
$\Alf \otimes \Alsspin$.
We express $\Alwhole$  as 
\begin{eqnarray*}
\label{eq:CAR6}
\Alwhole=\Matfour =\Mat \otimes \Mat \ \;{{\mbox {on}}} \ 
\Hil=\Hilf \otimes \Hils,
\end{eqnarray*}
 where  $\Mat\otimes \identitybf$
 is identified with 
the subsystem $\Alf$, and 
 $\identitybf \otimes \Mat$
 with  $\Alsspin$. Then 
 $\Alf$ acts on $\Hilf(\iso \C^{2})$ 
 while  $\Alsspin$ acts on
 $\Hils(\iso \C^{2})$;  they are algebraically  independent.

We note that 
\begin{eqnarray*}
\label{eq:Elsspin0}
\esspinoo  \equiv \bscr \bs,\quad 
\esspinot \equiv \bscr,\quad  %
\esspinto \equiv \bs,\quad %
\esspintt \equiv \bs \bscr
\end{eqnarray*}
  give a system of matrix units of $\Alsspin$.
It is easy to see 
\begin{eqnarray}
\label{eq:Elsspin1}
\esspinoo =\ascr \as,\quad   
\esspinot =\Uf\ascr,\quad  
\esspinto =\Uf\as,\quad 
\esspintt = \as \ascr.
\end{eqnarray}

Let $\xifo$  be an eigenvector of $\efoo$ in $\Hilf$
  belonging to the  eigenvalue 1, and let 
 $\xift\equiv\af \xifo$.
  Then  $\{ \xifo,\  \xift \}$ is  a CONS of $\Hilf$.
 Let $\xiso$  be an eigenvector of $\esspinoo$ in $\Hils$
  belonging to the  eigenvalue 1, and let 
 $\xist\equiv\bs \xiso$.
  Then  $\{ \xiso,\  \xist \}$ is  a CONS of $\Hils$.
We denote $\xifi \otimes  \xisj (\in \Hil )$  by $\xiij$. 
Then $\bigl\{ \xiij  \bigr\}_{i,j=1,2}$ is  a CONS of $\Hil$.
They are fixed once and for all, 
but their choice   is not essential for 
  all our  discussions.
\subsection{Pure States and  their Marginals}
\label{subsec:PUREG}
Let $\rho$ be an arbitrary pure state of $\Alwhole$.
It is represented by a unit vector   $\xi$ in $\Hil$,
 $\xi$ being fixed by $\rho$ up to a  phase factor. 
For $A \in \Alwhole$, its  expectation value is given by
\begin{eqnarray}
\label{eq:GVS0}   
\rho(A)=(A\xi, \, \xi)_{\Hil}.
\end{eqnarray}
  This $\xi$ can be decomposed by the CONS $\CONS$ as 
\begin{eqnarray}
\label{eq:GVS1}
\xi=\sum_{i,j=1,2} \ccij \xiij,
\end{eqnarray}
 where $\ccij \in \C$.
Due to  $\Vert  \xi \Vert=1$, 
\begin{eqnarray*}
\label{eq:GVS2}
\sum_{i,j} |\ccij|^{2}=1 
\end{eqnarray*}

We  calculate the density matrices of the reduced states of $\rho$
  to  subsystems of $\Alwhole$.
 $\rho|_{\Alf}$  and  $\rho|_{\Alsspin}$ 
The restrictions of $\rho$  to $\Alf$ and to 
$\Alsspin$ have the following 
density matrices:
\begin{eqnarray}
\label{eq:DENGaf}
\rho|_{\Alf}
= \left(
   \begin{array}{cc}
\vert \ccoo  \vert^{2} + \vert \ccot \vert^{2} & 
\ccoo \ccto^{\ast}
 +\ccot \cctt^{\ast}\\ 
\ccoo^{\ast}\ccto+ \ccot^{\ast} \cctt 
& \vert \ccto \vert^{2} + \vert \cctt \vert^{2}  
  \end{array}
 \right), 
\end{eqnarray}
\begin{eqnarray}
\label{eq:DENGsspin}
\rho|_{\Alsspin}
= \left(
   \begin{array}{cc}
\vert \ccoo  \vert^{2} + \vert \ccto \vert^{2} & 
\ccoo \ccot^{\ast}
 +\ccto \cctt^{\ast}\\ 
\ccoo^{\ast}\ccot+ \ccto^{\ast}\cctt  
&    \vert \ccot \vert^{2} + \vert \cctt \vert^{2}  
  \end{array}
 \right), 
\end{eqnarray}
 where the $(i, j)$ element  in  (\ref{eq:DENGaf})
 is given by  the expectation value of $\efji$ in the state $\rho$, 
 while the  $(i, j)$ element  in  (\ref{eq:DENGsspin})
 is given by that  of $\esspinji$.
 
Furthermore,  the density matrix of $\rho$ restricted to $\Als$
 is given by 
\begin{eqnarray}
\label{eq:DENGas}
\rho|_{\Als}
= \left(
   \begin{array}{cc}
\vert \ccoo  \vert^{2} + \vert \ccto \vert^{2} & 
\ccoo \ccot^{\ast}
 -\ccto \cctt^{\ast}\\ 
\ccoo^{\ast}\ccot- \ccto^{\ast}\cctt  
&    \vert \ccot \vert^{2} + \vert \cctt \vert^{2}  
  \end{array}
 \right), 
\end{eqnarray}
 where the $(i,j)$ element is  the expectation value
 of $\esji$ in the state $\rho$. 
\section{Failure  of Triangle Inequality of 
 von Neumann Entropy}
\label{sec:VIO}
Let $\Hil$ be an  $n$-dimensional Hilbert space and 
  $\LL(\Hil)$  be the  set of 
 linear operators on $\Hil$.
 The von Neumann  entropy of a state $\ome$ on $\LL(\Hil)$
 is given as usual by  
\begin{eqnarray*}
\label{eq:vonNeumann}
S(\omega) \equiv -\Tr\bigl( \Dome \log \Dome \bigr),
\end{eqnarray*}
 where $\Tr$ is  the  matrix trace  which takes the value $1$ 
 on each minimal projection  and  
$\Dome$  denotes the density matrix of $\ome$ with respect to $\Tr$. 
 It is well-known and easy to see  that  
 $\log n \ge S(\ome) \ge 0$, 
 $S(\ome)=\log n$ if and only if $\ome$ is the  (unique) 
tracial state $\tau(\cdot)=\frac{1}{n} \Tr(\cdot)$ of  $\LL(\Hil)$,
 and  $S(\ome)=0$ if and only if 
$\ome$ is a pure state of $\LL(\Hil)$. 

We introduce 
 the so-called triangle inequality of von Neumann entropy 
 in a  general situation.
  Let $\Al$ and $\Bl$ be a pair of subalgebras  
  of a finite-dimensional  $\cstar$-algebra $\Cl$.
 Let $\ome$ be a state of $\Cl$.
 Let  $\omeA$ and $\omeB$ be its 
restrictions 
  to $\Al$ and $\Bl$, respectively.
The following entropy inequality is referred to as the triangle inequality:
\begin{eqnarray}
\label{eq:TRIANGLES}
\vert  S(\omega_{A}) -  S(\omega_{B})\vert \le S(\omega).
\end{eqnarray}
For any finite-dimensional bipartite tensor-product  system  where 
 $\Al=\LL(\HilA)\otimes \identitybf_{B}$
 and   $\Bl=\identitybf_{A} \otimes \LL(\HilB)$, and $\Cl=\Al \otimes \Bl$, 
the above inequality holds for any  state $\ome$ of $\Cl$
 \cite{ARAKILIEB}. 

We  now give  a counterexample of the  triangle inequality
 for  our CAR system  where $\Alf$ and $\Als$
 are $\Al$ and $\Bl$, respectively, 
 and $\Alwhole$ is $\Cl$ in the above formula.

If we take $\ccij=\frac{1}{2}$ for all $i,j$ in (\ref{eq:GVS1}), 
then the (pure) state  of the total system  $\Alwhole$
 is uniquely determined  by (\ref{eq:GVS0}) 
 and will be denoted by $\hrho$.
By substituting $\frac{1}{2}$ into each $\ccij$ in 
(\ref{eq:DENGaf}), (\ref{eq:DENGsspin}), and (\ref{eq:DENGas}), 
 we have the following explicit formulae for the density matrices:
\begin{eqnarray}
\label{eq:DENhaf}
\hrho|_{\Alf}
= \left(
   \begin{array}{cc}
\frac{1}{2} & 
\frac{1}{2}\\ 
 \frac{1}{2}
& \frac{1}{2}
  \end{array}
 \right), 
\end{eqnarray}
\begin{eqnarray}
\label{eq:DENhsspin}
\hrho|_{\Alsspin}
= \left(
   \begin{array}{cc}
\frac{1}{2} & 
\frac{1}{2}\\ 
\frac{1}{2}
&    \frac{1}{2}
  \end{array}
 \right),
\end{eqnarray}
and 
 \begin{eqnarray}
\label{eq:DENhas}
\hrho|_{\Als}
= \left(
   \begin{array}{cc}
\frac{1}{2} & 
0 \\ 
0
& \frac{1}{2}
  \end{array}
 \right).
\end{eqnarray}
Therefore  $\hrho|_{\Alf}$ and $\hrho \vert_{\Alsspin}$
 are  pure states with   entropy 0 and
$\hrho |_{\Als}$ is a tracial state with the maximal
 entropy $\log 2$.
Hence  we obtain
\begin{eqnarray}
\label{eq:VIOTRI}
  \log2=\Bigl| S(\hrho|_{\Alf})-S(\hrho|_{\Als}) \Bigr| >S(\hrho)=0,
\end{eqnarray}
  yielding  a counterexample of the triangle inequality
 of von Neumann entropy.
\begin{rem}
\begin{rm}
\label{rem:SSA} 
The so-called strong subadditivity (SSA)  of von Neumann
 entropy 
(which was proved for tensor-product systems \cite{LIEBRUSKAI73})
is also shown  to  hold  for  our CAR systems \cite{ARAKIMORIYA}. 
This result  will be used in the proof of Proposition$\,$\ref{pro:PUREPURE}.
\end{rm}
\end{rem}
\section{Nonindependence  of   CAR Systems}
\label{sec:NONIND}
\subsection{States with Pure Marginal States}
\label{subsec:PUREPURE}
We  show  a formula  of states on  $\Alwhole$  
such that 
their  restrictions to $\Alf$ and  to $\Als$ are both pure states.
\begin{pro}
\label{pro:PUREPURE}
Let $\ome$ be  a  state of 
$\Alwhole$.
Suppose that its restrictions to $\Alf$ and 
to $\Als$ are both pure states. Then 
$\ome$ is a pure state of $\Alwhole$ and has the following   product property     over 
  $\Alf$ and  $\Alsspin$ $:$ 
\begin{eqnarray}
\label{eq:PUREPURE0}
\ome(AB)=\ome(A)\ome(B),
\end{eqnarray}
for every $A \in \Alf$ and $B \in \Alsspin$.
The restriction of $\ome$ to $\Alsspin$ is a pure state. 
\end{pro}
{\it Proof.} 
Let $\omef$ be the  restriction  of $\ome$ to $\Alf$ and 
$\omes$ be that to $\Als$.
By the assumption that $\omef$ and $\omes$ are pure states,
 both von Neumann entropies vanish:
\begin{eqnarray}
\label{eq:ENTzero}
S(\omef)=S(\omes)=0
\end{eqnarray}
 It follows from 
(\ref{eq:ENTzero}) and the subadditivity property of entropy
 for CAR systems proved in \cite{ARAKIMORIYA} that
\begin{eqnarray*}
 \label{eq:Strong}
S\bigl( \ome |_{\Alwhole} \bigr)  \leq S(\omef) +S(\omes)=0+0=0.
\end{eqnarray*}
 Thus  the positivity of entropy implies 
\begin{eqnarray*}
 \label{eq:ENTzero2}
S\bigl( \ome |_{\Alwhole} \bigr)  =0.
\end{eqnarray*}
 By  this  vanishing result of   entropy of $\ome$, we conclude  that 
$\ome$ is  a pure state of $\Alwhole$.
 Thus there exists a unique normalized 
vector $\etaome$ in $\Hil$ up to a  phase factor  satisfying 
\begin{eqnarray*}
\ome(A)=(A \etaome,\, \etaome)_{\Hil},\quad A \in \Alwhole.
\end{eqnarray*}

The product
 property (\ref{eq:PUREPURE0}) follows from the 
 lemma below. 
 By (\ref{eq:PUREPURE0}), the purity of $\ome$ implies 
 that of the restriction of $\ome$ to $\Alsspin$.
\proofend
\begin{lem}
\label{lem:TAKESAKI4.11}
Let  $\Hilf$
 and $\Hils$ be (arbitrary dimensional) Hilbert spaces, and 
 $\Hil=\Hilf \otimes \Hils$.
 If a state $\ome$
 of $\LL(\Hil)$ has a pure state 
restriction to $\LL(\Hilf)\otimes \identitybf_{\Hils}$, then $\ome$
 has  the following product property $:$ \\
\begin{eqnarray*}
\ome(AB)=\ome(A)\ome(B)
\end{eqnarray*} 
 for $A \in \LL(\Hilf)\otimes \identitybf_{\Hils}$ and  
$B \in \identitybf_{\Hilf} \otimes \LL(\Hils)$.
\end{lem}
This lemma is a well-known fact, see e.g., Lemma IV.4.11 of \cite{TAKESAKI1}.
\begin{rem}
\begin{rm}
\label{rem:IndeBETSU}
 For the $\ome$ in Proposition \ref{pro:PUREPURE},
  the same result holds for the pair
  $\Als$ and $\Alfspin$ 
  as that for $\Alf$ and $\Alsspin$. 
\end{rm}
\end{rem}

\begin{rem}
\begin{rm}
The purity of the both restrictions of  $\ome$
 to  $\Alf$ and $\Alsspin$
does not  imply the purity of that  to  $\Als$.
$\hrho$  in the preceding section    gives  an example; 
it is a product of a 
pure state
  on  $\Alf$ and  a pure state on  $\Alsspin$, but 
 has a  non-pure marginal    state  (tracial state) on  $\Als$.
\end{rm}
\end{rem}
\subsection{Showing Nonindependence}
\label{subsec:NONSEPproof}
We  recall the definition 
 of $\cstar$-independence \cite{HK64}.
\begin{df}
\label{df:Independence}
{\rm{
Let $\Al$ and $\Bl$ be  subalgebras of a $\cstar$-algebra $\Cl$.
The pair $(\Al,\, \Bl)$ (or $\Al$ and $\Bl$) are said to be
 $\cstar$-independent if and only if  for every state $\varpif$  of $\Al$
 and  every state $\varpis$  of $\Bl$ there exists a state $\varpi$
 of $\Cl$ such that $\varpi |_{\Al}=\varpif$ and 
    $\varpi |_{\Bl}=\varpis$. 
}}
\end{df}
%
We note that 
this  definition  does not exclude  noncommuting   
 pairs of algebras.
 In fact, there are  several examples 
   which  are   noncommuting $\cstar$-independent pairs,
 see, e.g., \cite{FS}, \cite{SUMMERS} and   references  therein.
 
We  now show  that a pair of $\cstar$-subalgebras 
 $\bigl( \Alf,\, \Als \bigr)$  of \,$\Alwhole$ 
  are not $\cstar$-independent. 
Let $\vrhof$ be an arbitrary pure state of $\Alf$ and $\vrhos$ be an arbitrary pure state of $\Als$. 
 Let us  assume that there exists a state $\vrho$ 
of $\Alwhole$ such that $\vrho |_{\Alf}=\vrhof$ and 
    $\vrho |_{\Als}=\vrhos$.
  Our aim is to  derive the inconsistency of this assumption
   for some pair of states $\vrhof$ and $\vrhos$  which leads to 
 the proof of the non-existence of such  $\vrho$. 

 Since both $\vrhof$ and $\vrhos$ are pure states,
 they are represented by the
 following density matrices 
with some positive numbers $\vtheta, \vthetas, \vp, \vps$ such that
$0 \leq \vtheta,\,  \vthetas  < 2 \pi$ 
and $0 \leq \vp,\, \vps \leq \frac{\pi}{2}$:
\begin{eqnarray}
\label{eq:Nvrhof}   %
\vrhof =\MatCARf{\vp}{\vtheta}, 
\end{eqnarray}
\begin{eqnarray}
\label{eq:KATEIs}
\vrhos =\MatCARf{\vps}{\vthetas}.
\end{eqnarray}
  Let us   denote   the restriction  
 of the state $\vrho$ to $\Alsspin$ by 
$\vrhospin$. 
 By Proposition \ref{pro:PUREPURE}, the assumption on this $\vrho$ 
 implies  that  $\vrhospin$ is a pure state and   
  $\vrho$ is  a pure  product state over $\Alf$ and $\Alsspin$
  in the form of $\vrhof \otimes \vrhospin$.

Since  $\vrhospin$ is a pure state, 
it is represented 
 by the following density matrix  with 
  $\vthetap \,(0 \leq \vthetap < 2 \pi)$ 
and $\vpp \,(0 \leq \vpp \leq \frac{\pi}{2})$
\begin{eqnarray}
\label{eq:Nvrhosspin}
\vrhospin=\MatCARf{\vpp}{\vthetap}.
\end{eqnarray}

By calculating the expectation values of 
 the matrix units of $\Als$ given by 
\begin{eqnarray*}
\label{eq:MATEles5}
\esoo  = \esspinoo ,\quad  \esot =\Uf \esspinot, \quad 
\esto =\Uf \esspinto,\quad  \estt=\esspintt
\end{eqnarray*}
for  $\vrho(= \vrhof \otimes \vrhospin)$,
 we   express 
the density matrix of $\vrhos$
 in terms of    $\vthetap,  \vpp$ and  $\vp$  
 as follows: 
\begin{eqnarray}
\label{eq:vrhosg}
\vrhos=\MatCARsg{\vpp}{\vp}{\vthetap},
\end{eqnarray}
 where $g(\vp) \equiv \cos^{2}(\vp)-\sin^{2}(\vp)$.

In order that (\ref{eq:KATEIs}) coincides with  (\ref{eq:vrhosg}),
 we must have
\begin{eqnarray*}
\cos^{2}(\vps)=\cos^{2}(\vpp),\quad 
\sin^{2}(\vps)=\sin^{2}(\vpp),\\
\cos^{2}(\vps)\sin^{2}(\vps)=g(\vp)^{2}\cos^{2}(\vpp)\sin^{2}(\vpp),
\end{eqnarray*}
and, hence, 
\begin{eqnarray*}
\bigl( g(\vp)^{2}-1 \bigr)\cos^{2}(\vps)\sin^{2}(\vps)=0.
\end{eqnarray*}
Then the one of the following must hold: \\ 
 (1)\,$\vps=0$,\quad  (2)\,$\vps=\frac{\pi}{2}$,\quad  (3)\,$g(\vp)^{2}=1$.

In the case (1) and (2), $\vrhos$ is diagonal.
In the case (3), either $\vp=0$ or $\vp=\frac{\pi}{2}$
 must hold, and hence $\vrhof$ is diagonal.
Therefore, if both $\vrhof$ and $\vrhos$ are not diagonal,
  there does not exist the state $\vrho$ whose restrictions to
 $\Alf$ and $\Als$ are $\vrhof$ and $\vrhos$, respectively.

In conclusion, we have shown the following theorem.
\begin{thm}
\label{thm:NONSEP}
$\Alf$ and $\Als$ are not $\cstar$-independent.
\end{thm}
\section{Half-sided Entanglement}
\label{sec:HALF}
\subsection{Definition of Entanglement Degree}
 \label{subsec:vNHALF} 
 We first give a simple  definition of  entanglement degree
 for rather general  situations  including  our CAR systems and 
finite-dimensional tensor-product systems as special cases.
\begin{df}
\label{df:defENTANGLE} 
{\rm{
Let $\Cl$ be a $\cstar$-algebra and $\Al$ 
 be a  finite-dimensional subalgebra  of $\Cl$.
Let $\ome$ be a state of \,$\Cl$.
 The quantum entanglement degree   of $\ome$
 on $\Al$ 
 is defined    by
\begin{eqnarray}
\label{eq:defENTANGLE}
E(\ome,\, \Al,\, \Cl) \equiv 
\inf_{\ome = \sum \lami \omei }
 \sum_{i} \lami    S(\omei|_{\Al}),
\end{eqnarray}
where the infimum is taken over all  convex decompositions of
 $\omega$
in the state space of $\Cl$. 
}}
\end{df}

By definition, for any pure state $\ome$ of $\Cl$,
\begin{eqnarray}
\label{eq:defENTANGLE2}
E(\ome,\, \Al,\, \Cl) =
 S(\ome|_{\Al}).
\end{eqnarray}
If $\Al$ and $\Bl$ are  finite-dimensional  matrix algebras
  and 
 $\Cl=\Al \otimes \Bl$, then  
\begin{eqnarray}
\label{eq:pureSYM}
\Bigl|S(\ome|_{\Al})  -S(\ome|_{\Bl})\Bigr|=0 
\end{eqnarray}
 for any pure state $\ome$ of $\Cl$. Hence, 
\begin{eqnarray}
\label{eq:Formation1}
E(\ome,\, \Al,\, \Cl)= 
\inf_{\ome = \sum \lami \omei} \lami S(\omei |_{\Al} )
 = \inf_{\ome = \sum \lami \omei} \lami S(\omei |_{\Bl} )
=E(\ome,\, \Bl,\, \Cl).
\end{eqnarray}   
 for any state $\ome$ of $\Cl$. 
 For this  case, entanglement $E$ is symmetric in $\Al$
 and $\Bl$. However, it is not true in general 
as  we  will see.
\begin{rem}
\begin{rm} 
\label{rem:EXCUSE}
Different  entanglement degrees 
are known  for  finite-dimensional tensor-product systems.
For this case,  uniqueness theorems of  entanglement degrees 
  on  pure states have  been shown,  asserting  that 
all possible entanglement degrees satisfying some 
 basic postulates  are equal to
   the von Neumann entropy of the marginal states (of the 
pure states).
 (For details, see e.g. \cite{DONALD3}, \cite{RUD}, \cite{HHH.PRL84}  
 and references therein.) 

We give  our entanglement degree (\ref{eq:defENTANGLE})
 as a straightforward generalization of ``entanglement of formation"
 (\ref{eq:Formation1}) which was  defined 
  for finite-dimensional tensor-product systems
 in \cite{BEN.PRA54}.
 We leave  
its justification as a natural entanglement degree 
 for  the CAR case as an open problem,
although it is a crucial matter. 
\end{rm} 
\end{rem}
\subsection{Asymmetry of Entanglement}
We   
 introduce a new notion named `half-sided entanglement'
 in this Section.
The  contrast 
between CAR systems  
and tensor-product systems as seen in  (\ref{eq:VIOTRI}) and 
(\ref{eq:pureSYM})
 leads us 
 to  an  intuitive understanding:\ 
asymmetry of quantum entanglement  is caused by
 the   nonindependence of the pairs of subsystems. 
We  now give the following definitions.
\begin{df}
\label{df:EBS} 
{\rm{
Let $\Cl$ be a $\cstar$-algebra and $\Al$ and $\Bl$
 be (a pair of) finite-dimensional subalgebras of $\Cl$.
Let $\ome$ be a state of \,$\Cl$.
 The degree of $S$-asymmetric entanglement  of $\ome$
 between $\Al$ and $\Bl$ 
 is given   by
\begin{eqnarray}
\label{eq:EBS}
\EBS(\ome,\, \Al, \, \Bl, \, \Cl) \equiv 
\inf_{\ome = \sum \lami \omei }
 \sum_{i} \lami     \Bigl| S(\omei|_{\Al})  -S(\omei|_{\Bl}) \Bigr|,
\end{eqnarray}
where the infimum is taken over all  convex decompositions of
 $\omega$
in the state space of $\Cl$. 
}}
\end{df}
 \begin{df}
\label{df:EBS2}
{\rm{
If  $\EBS(\ome,\, \Al,\, \Bl,\, \Cl)$ is nonzero,  
$\ome$ is said to be an $S$-asymmetrically  entangled state 
 with respect to $(\Al,\, \Bl)$. 
 
 Let  $\bigl\{ \lami, \, \omei\bigr\}$ 
 be  a state-decomposition  of \,$\ome$ attaining 
 $\EBS(\ome,\, \Al,\, \Bl,\, \Cl)$,
 that is,  
$\ome =\sum_{i} \lami \omei$ 
 and 
\begin{eqnarray}
\label{eq:EBSattain}
\EBS(\omega,\, \Al,\, \Bl,\, \Cl) = 
 \sum_{i} \lami \Bigl| S(\omei|_{\Al})  -S(\omei|_{\Bl}) \Bigr|.
\end{eqnarray}
 If each $\omei|_{\Al}$ is a pure state, and hence  
$\sum_{i} \lami  S(\omei|_{\Al})=0 $,
 then $\ome$ is said to have    $S$-half-sided entanglement 
  $\EBS(\ome,\, \Al,\, \Bl,\, \Cl) $
  on $\Bl$ with respect to  $(\Al,\, \Bl)$. 

If  $\ome$ takes the maximal value of 
 $\EBS(\cdot\,,\, \Al,\, \Bl,\, \Cl)$ when it exists,  
that is, 
\begin{eqnarray*}
\EBS(\ome,\, \Al,\, \Bl,\, \Cl)
&=& \sup_{  
{\omep :\;  {\mbox{\scriptsize{state of }}}  \;\Cl   }  
 } \EBS(\omep,\, \Al,\, \Bl,\, \Cl),
\end{eqnarray*}
 then $\ome$  is said to have   maximal $S$-asymmetric entanglement.

 If $\ome$   has
    maximal   $S$-asymmetric  entanglement  and at the same time 
  is  $S$-half-sided entangled, it is said to be a 
maximal  $S$-half-sided entangled state.
}}
\end{df}

`$S$-' in the above  definitions 
  refers to the von Neumann entropy. 
Obviously,
\begin{eqnarray}
\label{eq:EBSsym}
\EBS(\ome,\, \Al,\, \Bl,\, \Cl)=\EBS(\ome,\, \Bl,\, \Al,\, \Cl),
\end{eqnarray}
  for any state $\ome$ of $\Cl$,  and 
\begin{eqnarray}
\label{eq:EBSpure}
\EBS(\ome,\, \Al, \, \Bl, \, \Cl)
&=&\Bigl| S(\ome|_{\Al})  -S(\ome|_{\Bl}) \Bigr|, \nonum \\
&=& \Bigl|   E(\ome,\, \Al,\, \Cl)-E(\ome,\, \Bl,\, \Cl) \Bigr|,
\end{eqnarray}
  for any pure state $\ome$ of $\Cl$.
Since we take the infimum over all the possible convex decompositions of $\ome$
 in the state space $\Cl$,
 $\EBS(\omega,\, \Al,\, \Bl,\, \Cl)$  is a convex function of $\ome$.

\begin{rem}
\begin{rm} 
\label{rem:EBSTENkill}
Let $\Cl=\Al \otimes \Bl$ and both $\Al$
 and $\Bl$ be  finite-dimensional matrix algebras.
 It follows  from (\ref{eq:pureSYM}) and  (\ref{eq:EBSpure}) that  
\begin{eqnarray}
\label{eq:EBSTENkill}
\EBS(\ome,\, \Al,\, \Bl,\, \Cl) =0, 
\end{eqnarray} 
 for any
  state $\ome$ of $\Cl$.
\end{rm}
\end{rem}
\begin{rem}
\begin{rm}
\label{rem:3SYSTEM}
When $\Cl \supset \Al \otimes \Bl$,  
 (\ref{eq:EBSTENkill}) does not  hold in general.
 A  counterexample  is given as follows.
 Let $\Al$ and $\Bl$ be finite-dimensional $\cstar$-algebras.
 Consider $\Cl=\Mat \otimes \Al \otimes \Bl$.
 Let $\omef$ be a pure state on $\Mat \otimes \Al$
 and $\omes$ be a pure state on $\Bl$.
 Let us give the (total) state $\ome$ 
on  $\Cl$
as the  product state 
 of $\omef$ and $\omes$.
$\ome$ has  the only trivial decomposition over $\Cl$ because of its purity.
$S(\ome|_{\Bl})$ vanishes, but $S(\ome|_{\Al})\ne 0$
 unless $\omef$ has the product property  over  $\Mat$ and $\Al$.

\end{rm}
\end{rem}
\section{Entanglement in CAR Systems}
\label{sec:CARE}
\subsection{$S$-asymmetric Entanglement in CAR Systems}
\label{subsec:EBSCAR}
We calculate 
 $\EBS(\cdot\, ,\, \Alf,\, \Als,\, \Alwhole)$,
 $E(\cdot\, ,\, \Alf,\, \Alwhole)$ and 
 $E(\cdot\, ,\, \Als,\, \Alwhole)$
 for  some  states
 of   the following specific form. 
 The state $\vrho$ of $\Alwhole$ 
to be considered is    a pure  state of 
 the form 
\begin{eqnarray}
\label{eq:vrho}
\vrho =\vrhof \otimes \vrhospin,
\end{eqnarray}
 where $\vrhof$ is a pure state
 of $\Alf$ given  by the density matrix (\ref{eq:Nvrhof}),
 and  $\vrhospin$   is a pure state
 of $\Alsspin$ given by the density matrix  (\ref{eq:Nvrhosspin}).
By our choice, $\vrho$  is a vector state 
 whose  representative vector
$\etavrho \in \Hil$  has the  product form
\begin{eqnarray*}
\label{eq:etavrho1}
\etavrho = \etafvrho \otimes \etasvrho,
\end{eqnarray*}
 where  $\etafvrho \in \Hilf$ and $\etasvrho \in \Hils$
can be taken  as  
\begin{eqnarray*}
\label{eq:etavrho2}
\etafvrho &\equiv&  e^{i \vtheta} \cos (\vp) \xifo
   +  \sin (\vp) \xift,  \nonumber \\
\etasvrho &\equiv& e^{i \vthetap}
 \cos (\vpp) \xiso
   + \sin (\vpp) \xist.
\end{eqnarray*}

Let $H(\cdot,\, \cdot)$ be an entropy function
 given by the following formula,
\begin{eqnarray*}
\label{eq:Hfunction}%
H(a,\,b) \equiv - a \log a - b \log b,
\end{eqnarray*}
 for two  positive numbers $a,\,b$. 
Let $\vrhos$ be a restriction of the state $\vrho$
 to $\Als$.  Its density matrix  is given
 as (\ref{eq:vrhosg}). The eigenvalues $\vrhos^{\pm}(\vpp,\,\vp)$
 of $\vrhos$ are  given by
\begin{eqnarray*}
\label{eq:EIGENvrhos}
\frac{1 \pm \sqrt{1-4 
 \Bigl( 1-\bigr\{ g(\vp) \bigr\}^{2} \Bigr) \cdot
 \cos^{2}(\vpp)\sin^{2}(\vpp)
                  } 
      }{2}.
\end{eqnarray*}
 Since $\vrhof$ is a pure state of $\Alf$, we have
\begin{eqnarray*}
\label{eq:EBS4}
 S(\vrho|_{\Alf})=S(\vrhof)=0.
\end{eqnarray*}
Hence 
\begin{eqnarray}
\label{eq:Evrhof}
E(\vrho\, ,\, \Alf,\, \Alwhole)=0.
\end{eqnarray}

We have also
\begin{eqnarray*}
\label{eq:EBS5}
S(\vrho|_{\Als})=S(\vrhos) 
   =H\bigl( \Evrhosp,\,\Evrhosm \bigr).
\end{eqnarray*}
Hence 
\begin{eqnarray}
\label{eq:Evrhos}
E(\vrho\, ,\,  \Als,\, \Alwhole)= H\bigl( \Evrhosp,\,\Evrhosm \bigr).
\end{eqnarray}
Thus we obtain
\begin{eqnarray} 
\label{eq:EBSvrho}
   \EBS(\vrho,\, \Alf,\, \Als,\, \Alwhole)
  =H\bigl( \Evrhosp,\,\Evrhosm \bigr).
\end{eqnarray} 

 For any fixed $\vp$, $H\bigl( \Evrhosp,\,\Evrhosm \bigr)$
 increases with $\vpp$ from $\vpp=0$ until $\vpp=\frac{\pi}{4}$,
 and then decreases from $\vpp=\frac{\pi}{4}$ until $\vpp=\frac{\pi}{2}$.
Unless $\vp =0$ or $\vp= \frac{\pi}{2}$, 
 namely unless $\{ g(\vp) \}^{2}=1$, 
it first increases strictly
until $\vpp=\frac{\pi}{4}$ and then decreases  strictly.

On the other hand, 
for any fixed $\vpp$, $H\bigl( \Evrhosp,\,\Evrhosm \bigr)$
 increases with $\vp$ from $\vp=0$ until $\vp=\frac{\pi}{4}$,
 and then decreases from $\vp=\frac{\pi}{4}$ until $\vp=\frac{\pi}{2}$.
Unless $\vpp =0$ or $\vpp= \frac{\pi}{2}$, it first increases strictly
until $\vp=\frac{\pi}{4}$ and then decreases  strictly.
Unless $\vpp =0$ or $\vpp= \frac{\pi}{2}$, and  at the same time 
unless 
   $\vp =0$ or $\vp= \frac{\pi}{2}$, 
$\vrho$ has  strictly positive  $S$-half-sided entanglement  
$H\bigl( \Evrhosp,\,\Evrhosm \bigr)$ on $\Als$ with respect 
 to $(\Alf,\,\Als)$.

 If $\vpp=\vp=\frac{\pi}{4}$, then 
$\vrhos$ is a tracial state. Hence 
   $\EBS(\vrho,\, \Alf,\, \Als,\, \Alwhole)$ takes the maximal value
$\log 2$. Therefore, this $\vrho$ 
  is a maximal  $S$-half-sided entangled state 
 on $\Als$ with respect  to $(\Alf,\,\Als)$.
We have  shown the following.
\begin{thm} 
\label{thm:S}
For any positive number $x \in [\,0\,,\; \log 2\,]$, 
 there exists an \,$S$-half-sided entangled state of $\Alwhole$
  for the pair $(\Alf,\, \Als)$
 with its degree of S-asymmetric entanglement $x$.
\end{thm}
\begin{rem}
\begin{rm}
\label{rem:IIWAKE}
We   may add  a remark  on  our terminology 
 `half-sided' entanglement to avoid a possible 
 misunderstanding. 
 Entanglement  is not something 
  which can be localized  or  concentrated  physically 
in a  single local system (half-sided system).
It refers to nonlocal correlations
shared by  subsystems in an entire  system.
 Half-sided entanglement will  describe 
  those asymmetric features
 of entanglement shared by nonindependent pairs, 
which cannot  be observed 
 in  any algebraically independent  pairs.
\end{rm}
\end{rem}
\subsection{Operations on the Half-sided System}
  It is natural to expect  some 
{\it{operational nonlocality}}  accompanies 
 with  nonindependent systems.
We  show  how quantum entanglement between $\Alf$ and 
$\Als$
 will be effected 
 by   operations done in  the half-sided system $\Alf$.

By local automorphisms of $\Alf$, we mean 
 the automorphisms in the form of 
$\alpha_{1} \otimes \identitybf_{\Alsspin}$,
the  tensor product of some  automorphism $\alpha_{1}$
 of $\Alf$ and the identity map of $\Alsspin$.
 In general,   $\Als$ is not  invariant as a set under 
a local automorphism of $\Alf$.
We shall see that  local automorphisms of $\Alf$ 
 can  change  the entanglement 
degree between $\Alf$ and $\Als$.

We consider the  set of states of $\Alwhole$ 
in the form of 
  $\vrho =\vrhof \otimes \vrhospin$,
 where $\vrhof$ is  given  by (\ref{eq:Nvrhof}),
 while $\vrhospin$   is given by   (\ref{eq:Nvrhosspin}).
  Fixing    the parameters
 $\vtheta$ and $\vp$  of  (\ref{eq:Nvrhof})
  and  $\vthetap$ and $\vpp$  of (\ref{eq:Nvrhosspin}), we have 
 an initial state 
 $\vrho_{\circ}$ of $\Cl$.
 By acting 
  local automorphisms of $\Alf$, we can transform this $\vrho_{\circ}$
 to any state  in the  form  $\vrho =\vrhof \otimes \vrhospin$   
 where  
  $\vrhof$  is given by 
  (\ref{eq:Nvrhof})
  with arbitrary $\vtheta$ and $\vp$ 
  while  $\vrhospin$ is kept 
 fixed as $\vrho_{\circ}|_{\Alsspin}$. 
Just recalling 
(\ref{eq:Evrhos}) and (\ref{eq:EBSvrho}), 
 we have 
\begin{eqnarray*}   
 E(\vrho,\, \Als,\, \Alwhole)=
 \EBS(\vrho,\, \Alf,\, \Als,\, \Alwhole)
=H\bigl( \Evrhosp,\,\Evrhosm \bigr).
\end{eqnarray*}
The above (equivalent) functions vary  with $\vp$.
Consequently, we have shown that 
 the entanglement degree of $\Als$ and the $S$-asymmetric entanglement
 degree of $\vrho$ on $\Als$ with respect to $(\Alf,\,\Als)$ can   
  change  under the 
 local automorphisms  induced by  $\Alf$.
\begin{rem}
\begin{rm}
 Invariance under local automorphisms and the nonincreasing property
under local operations are considered as 
basic  desiderata for  natural
 entanglement degrees. As for the tensor-product case, see e.g.
 \cite{DONALD3}, \cite{HHH.PRL84} for details.
 For CAR systems, effects induced by 
 {{\it half-sided}} operations
(operations made solely by 
the subsystem in a half-sided 
region)  are nonlocal, 
and hence our results in this subsection do not conflict 
 with those desiderata.
\end{rm}
\end{rem}

\noindent{\bf{Acknowledgements}}. 
The author acknowledges  the JSPS postdoctoral fellowship.
 He also thanks H.Araki and H.Narnhofer for their 
 comments on this work.
%
%

\end{document}